# ON THE EVALUATION OF GMSK SCHEME WITH ECC TECHNIQUES IN WIRELESS SENSOR NETWORK

Rajoua Anane , Kosai Raoof  and  Ridha Bouallegue

Laboratory of Acoustics at University of Maine, LAUM UMR CNRS n° 6613, France,
Innov'COM, University of Carthage, Tunisia,

*ABSTRACT*

*Wireless sensor nodes are powered by batteries, for which replacement is very difficult. That is why, optimization of energy consumption is a major objective in the area of wireless sensor networks (WSNs).On the other hand, noisy channel has a prominent influence on the reliability of data transmission. Therefore, an energy efficient transmission strategy should be considered on the communication process of wireless nodes in order to obtain optimal energy network consumption. Indeed, the choice of suitable modulation format with the proper Error Correcting code (ECC) played a great responsibility to obtain better energy conservation.In this work, we aim to evaluate the performance analysis of Gaussian Minimum Shift Keying (GMSK) modulation with several combinations of coding strategies using various analysis metrics such as Bit Error Rate (BER), energy consumption.Through extensive simulation, we disclose that he gain achieved through GMSK modulation with suitable channel coding mechanism is promising to obtain reliable communication and energy conservation in WSN.*

## 1. INTRODUCTION

In recent years, WSN have received considerable attention and it has become a prominent part of everyday life thanks to their special features (easy deployment, low cost, small size, multifunctional). This type of network has been used to monitor and report data in several applications such as, agriculture, health care, military, industry, tracking environmental pollution levels, and homeland security [2].

However, wireless sensor nodes are seriously constrained by energy supply. Therefore, how to enhance the network lifetime is a prominent and challenging issue.

In order to reduce energy consumption, wireless sensor employs some optimization strategy like switching from active mode to standby or sleep mode when there is no data to send. Indeed, the energy consumed during communicating process is significantly higher compared to the energy consumed during listening or processing. Hence, when the node has no information to transmit, it switches the standby/sleep mode to conserve its energy [1] [3].

 This strategy helps to reduce energy wastage, but it is not enough to solve the problem of energy constrained in WSN. Thereby, to extend the lifespan of wireless sensor network energy efficient transmission approach should be adopted at the transceiver of wireless sensor. Therefore, in the sensor node communication process, the choice of suitable modulation scheme and ECC technique plays an important role to obtain better energy conservation [4].

In the recent literature, the effect of various modulation schemes with different combinations of coding method has been addressed, to achieve a better overall system performance.

Three frequently used modulation techniques namely MQAM, MPSK and MFSK are evaluated without coding channel in [8]. Energy optimization issues have been analysed in point-to-point wireless communication under AWGN channel. This study discloses that the energy performance of MQAM and MPSK scheme is similar for all transmission ranges. But MFSK modulation is more energy efficient just in large distances.

Authors in [9] present a comparative analysis of uncoded MPSK, MQAM and MFSK schemes for short distance (less than 10 meters) and through both AWGN and Rayleigh channel. They conclude that MQAM scheme is more energy efficient than the other modulations.

Moreover, Performance evaluation of uncoded 16QAM; 64QAm, 16DPSK and 64DPSK modulation under AWGN, Rayleigh and Rician channel models is compared in [10] based on BER versus Signal-to-Noise Ratio (SNR). The authors observed that 16-QAM is performing better than 64-QAM.

In [11] the performance of binary-BCH codes and Reed Solomon codes are studied and analysed based on Bit Error Rate by transmitting randomly generated data using Phase Shift Keying (BPSK) modulation and AWGN channel condition. This work concludes that binary-BCH codes with ASIC implementation can be a suitable choice for wireless sensor networks.

In our earlier study [4], the performance of energy consumption referring to Minimum Shift Keying (MSK) modulation with suitable error control codes approach is analytically analyzed and simulated through a Gaussian channel (AWGN).

This paper builds on our earlier work in [4] [5] and analyses the performance of Gaussian Minimum Shift Keying modulation with coding approach over AWGN channel condition in order to find an energy efficient wireless sensor transmission strategy.

## 2. OUR CONTRIBUTION

As we have previously mentioned, performance analysis of different modulation techniques was studied in our earlier work [5]. It investigated the best modulation strategy to minimize the total energy consumption required to transmit a given number of bits. Indeed, the performance of several modulation formats including M-ary Quadrature amplitude modulation (MQAM), M-ary Phase-Shift Keying (MPSK), M-ary frequency-shift keying (MFSK) and minimum-shift keying (MSK) modulation are analysed based on energy consumption, communication distance, transmission time and modulation rate. In that study, we deduced that MSK modulation becomes more advantageous than its counterparts from the point of view of energy consumption.

Then, our study in [4] extends the obtain results and focus on improving the performance of MSK scheme with coding mechanism through AWGN channel. In this work, the performance analysis is assessed in terms of probability of Bit Error Rate (BER) for various error control codes and energy consumption. A deep campaign of extensive simulations discloses that the combination MSK modulation and Reed Solomon method provides a significant enhancement of sensor power consumption.

In this paper we intend to evaluate the performance of another popular modulation scheme called Gaussian Minimum Shift Keying (GMSK). This famous technique is based on MSK and frequently used in a variety of wireless networks like CDPD (cellular digital packet data), Global System for Mobile communications (GSM) system, Multiple-Input Multiple-Output (MIMO) approach and Code division multiple access (CDMA) techniques. GMSK scheme is a result of the attempts to improve the MSK power spectrum. It utilizes a low pass filter to provide data transmission with efficient spectrum usage [13].

Hence, our goal is to appraise the performance of GMSK modulation for both uncoded and coded systems through different computer simulations performed under MATLAB/Simulink software, in order to improve the overall energy conservation within wireless sensors.

This paper is organized as follows: The next section presents the WSN energy model adopted in our work. Section 4 offers a detailed performance analysis of GMSK modulation. An overview of channel coding is presented in section 5. The evaluation for GMSK scheme with coding strategy is developed in Section 6. Simulation results are discussed in Section 7. Finally, the last section concludes our work.

## 3. WIRELESS SENSOR NETWORK ENERGY MODEL

In this work, we performed the transmitter and receiver hardware model as introduced in our earlier work [4] [5] and shown in figure 1.

The transmitter block contains the following components:
- Digital to Analog converter (DAC),
- Filter,
- Frequency synthesizer,
- Mixer
- Power amplifier (PA).

The receiver part is composed of:
- Filter,
- Low noise amplifier (LNA),
- Frequency synthesizer,
- Mixer,
- Intermediate frequency amplifier (IFA),
- Analog to digital converter (ADC).

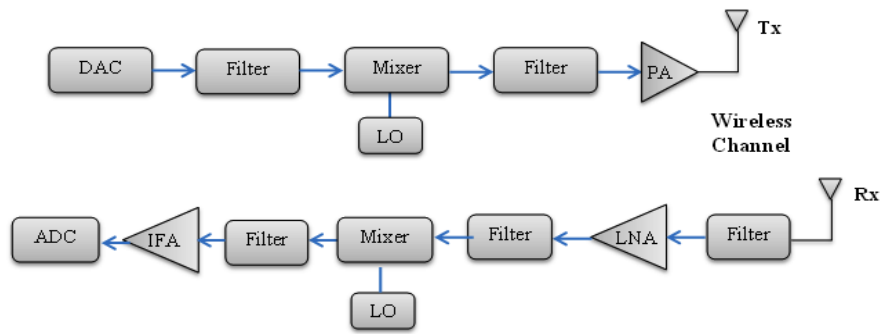

Figure 1. Transmitter and receiver hardware model

International Journal of Wireless & Mobile Networks (IJWMN) Vol. 7, No. 2, April 2015The energy consumed by both the transmitter and the receiver blocks was evaluated for calculating the total energy consumption in the network.

Case of frequency modulation schemes (MSK and GMSK), power consumption of both the DAC and the mixer at the transmitter were not taken into account in the calculation of the total power consumption [9] [14].

The transceiver block of a wireless sensor operates in three modes: the nodes works in the active mode when there is information to communicate, so all these circuits are active.
When there is no data to transmit the circuits switch to standby state. This technique contributes to energy saving. The power consumption in this mode is negligible. Knowing that switching from standby mode to active mode, the energy overhead caused by start-up transients is also significant and must be included in the calculation. This temporary state called transient mode which is used to initiate the frequency synthesizer at the local oscillator.

To sum up, for a specific hardware the energy consumed during the transient phase is considered constant but in a sleep mode we can assume that it is equal to zero.

Based on the above assumptions, the transmission period $T$ can be written as:

$$T = T_{start} + T_{on\_time} + T_{stby} \quad (1)$$

Where:
- $T_{start}$ is the time of the transient mode.
- $T_{on\text{-}time}$ represents the time spent to transmit L bits.
- $T_{stby}$ is the duration of the standby mode.

Power consumption associated to the described modes is denoted as:
- $P_{start}$: Power consumed for mode changing.
- $P_{on\text{-}time}$: Power consumed for transition
- $P_{stby}$: Power consumed in the standby state (assumed to be null for simplification)

More, we can derive the equation of the total energy consumed as follows:

$$\begin{aligned} E_{total} &= P_{on\_time} T_{on\_time} + P_{start} T_{start} \\ &= (P_{tx} + P_{tx-circuit} + P_{rx-circuit} + P_{PA}) T_{on\_time} + (P_{tx-circuit} + P_{rx-circuit}) T_{start} \end{aligned} \quad (2)$$

Where:
- $P_{tx}$ represents the power of data transmission.
- $P_{tx\text{-}circuit}$ and $P_{rx\text{-}circuit}$ are respectively circuit powers for transmitter and receiver without considering the amplifier.

We denote:
- $P_x$ is the power consumption of device $x$.

Expressing each term:

$$P_{tx-circuit} = P_{filt} + P_{syn} \quad (3)$$

$$P_{rx-circuit} = P_{ADC} + P_{filt} + P_{mixer} + P_{syn} + P_{LNA} + P_{IFA} \quad (4)$$

The power of the amplifier is expressed as [14]:

$$P_{PA} = \beta P_{tx} = \left(\frac{\xi}{\eta} - 1\right) P_{tx} \quad (5)$$

Where:
- $\eta$ represents the drain efficiency of the amplifier.
- $\xi$ represents the peak to average ratio. It depends on the modulation technique. $\xi = 1$ for frequency modulations i.e. GMSK [14]:

The total energy expression for GMSK modulation techniques are derived as:

$$E_{total-GMSK} = (1+\beta)P_{tx}T_{on-time} + (P_{ADC} + 2P_{filt} + P_{mixer} + 2P_{syn} + P_{LNA} + P_{IFA})T_{on-time} + 2P_{syn}T_{start} \quad (6)$$

## 4. GMSK PERFORMANCE ANALYSIS

GMSK modulation is considered as a filtered version of Minimum Shift Keying scheme since it implements a Gaussian low-pass filter and introduces a new variable called BT, where B represents the 3dB point of the Gaussian filter and the variable T is the bit period of the transmission process [13].

A bound on the probability of error for GMSK is expressed as [1]:

$$P_e = Q\left\{\sqrt{\frac{2\alpha E_b}{N_0}}\right\} \approx \frac{1}{2}erfc(\sqrt{\alpha SNR}) \quad (7)$$

Therefore, we can deduce:

$$P_e \approx e^{-\alpha SNR} \quad (8)$$

At the receiver side, the expression of the energy per information bit is written as:

$$E_{rxb} = N_0 N_f SNR \approx \frac{2}{\alpha}\sigma^2 N_f \ln(1/P_e) \quad (9)$$

In the output of the transmitter, the power of the signal is calculated by the equation of $K^{th}$ path loss model [15]. We can state that:

$$P_{tx} = P_{rx}G_d \quad (10)$$

Where $G_d = G_1 d^k M_1$ represents the power gain factor with:
- $G_1$: the gain factor at distance equal to $1m$,
- $M1$: the link margin
- $d$: the distance that separate two communicating nodes in (meters).
- $k$: The exponent order is between 2 and 4, in this paper k=3 is selected.

By following the same process used in our previous work [4] and knowing that $P_{rx} = (L E_{rxb} / T_{on-time})$ and $E_{tx} = P_{tx}T_{on-time}$, we obtain the expression of the transmission energy:

$$E_{tx-GMSK} = \frac{2}{\alpha}N_f \sigma^2 \ln(1/P_e)G_d L \quad (11)$$

Using (2), (6) and (11), we deduce the GMSK total energy consumption:

$$E_{total-GMSK} = \frac{2}{\alpha}(1+\beta) N_f \sigma^2 \ln(1/P_e)G_d L + P_{circuit}T_{on-time} + 2P_{syn}T_{start} \quad (12)$$

As well, the energy consumption per information bit is calculated as follows:

$$E_{inf\ bit-GMSK} = \frac{E_{total\_GMSK}}{L} \quad (13)$$

Correspondingly, the total energy consumption per information bit for GMSK is:

$$E_{inf\ bit-GMSK} = \frac{2}{\alpha}(1+\beta) N_f \sigma^2 \ln(1/P_e)G_d + \frac{P_{circuit}T_{on-time}}{L} + \frac{2P_{syn}T_{start}}{L} \quad (14)$$

## 5. OVERVIEW OF ECCS

Before starting the performance study of GMSK coded systems, it is essential to present an overview of coding channel techniques.

During the transmission process, the transmission data passes through a noisy channel. Due to this random noise, errors can be introduced in the received signal. These errors can be detected and sometimes corrected using coding methods. Therefore, Channel coding helps to minimize the effects of a noisy transmission channel.

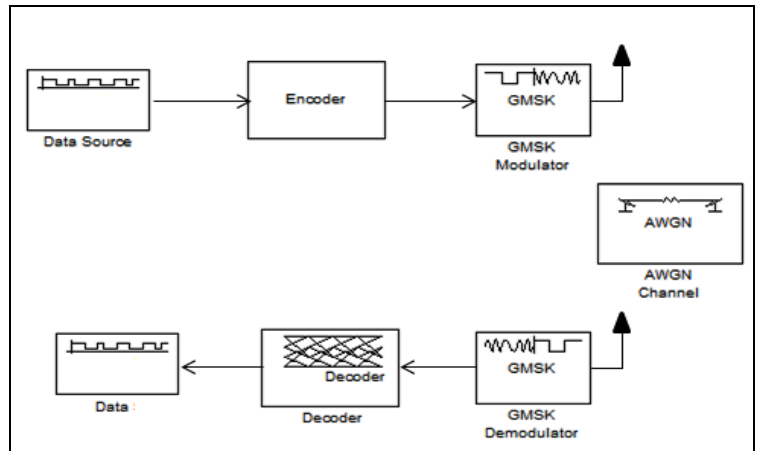

Figure 2. GMSK system model with channel coding

Generally, Channel error control codes are classified into two main categories [16]:
- Error detection codes: when an error is detected, a negative acknowledge is sent from the destination node to the source node, which request retransmission of data.
- Forward error correction codes (FEC): the errors are detected and corrected by ECC process at the receiver side.

Coding methods introduce redundancy into a data sequence and these extra bits are utilized to detect noisy received bits at the receiver sensor. For a block of length k bits, (n-k) check bits are added [17]. This means that the length of the code-word at the output of encoder block is n bits.

The ECC techniques can be classified in to two categories [18]:

- Convolutional Code takes a stream of data bits and converts it into a stream of transmitted bits, using a shift register bank. This type of coding often uses a Viterbi algorithm for decoding process.
- Linear Block Code encodes data in blocks with fixed lengths. There is a vast number of examples for block codes such as: Hamming, Golay, Reed Solomon, BCH. A linear code of length $n$, dimension $k$, and minimum distance $d$ is called an $[n,k,d_{min}]$ code with a rate $R=k/n$. $d_{min}$ represents the minimum number of different bites between any of the code-words and can be calculated as follows: $d_{min}= n-k+1$.

Applying a coding approach over a noisy channel can improve the performance of BER for the same value of Signal-to-Noise Ratio compared to a communication without coding process, or can also give the same bit error rate for a lower SNR. There is a trade-off between decoder complexity and coding gain. Indeed, to obtain higher gain we must use a long code that demands more complex decoder and subsequently high energy consumption.

## 6. EVALUATION OF GMSK SCHEME WITH CODING CHANNEL

It is clear that error correcting code permits minimization of the required transmitter signal energy due to its coding gain, but not all coding methods are suitable for the wireless network.

In this work, we have investigated different ECC performance analyses in terms of their Bit-Error-Rate as a function of Signal-to-Noise Ratio curves with GMSK scheme under AWGN channel model. Through extensive simulation we selected the optimal code among various coding techniques such as Golay, Reed Solomon (RS) and Convolutional codes.

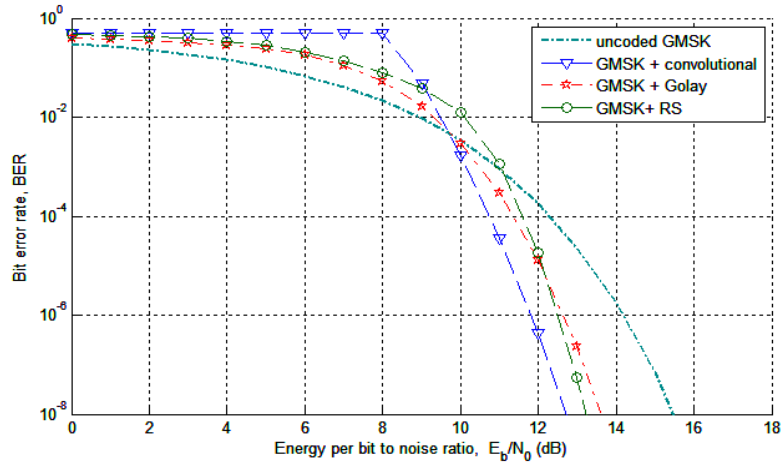

Figure 3. BER performance of GMSK under AWGN channel

Figure 3 illustrates the comparison between uncoded and coded GMSK scheme with several coding techniques. From these curves, we observe that for BER more than $10^{-2}$, coding methods performs worse than the uncoded system. But for BER lesser than $10^{-3}$, the coded system becomes more advantageous. We deduce also, that Convolutional code and Golay code achieve better characteristic than RS code.

It's true that Convolution code provides the highest gain but it requires high power consumption due to its encoding and decoding complexity. As a result, it is not suitable for energy constraint wireless sensors.

Hence, GMSK modulation with Golay code strategy is considered as the suitable combination. Thereby, we choose the Golay technique, which have the second highest gain, as optimized coding codes for the rest of our simulations.

On other hand, The Golay code permits to minimize the transmitter signal energy due to its coding gain $G_{code}$. Therefore, the required transmission energy is minimized by $G_{code}$ at a price of increased transmission time that will be expressed as follows:

$$T_{on-time-code} = T_{on-time}/R \quad (15)$$

From equation (2), the total energy consumption for coded system is deduced as follows:

$$E_{total} = (P_{enc} + P_{on-time} + P_{dec})T_{on-time-code} + P_{start}T_{start}$$
$$= (P_{enc} + P_{tx} + P_{tx-circuit} + P_{rx-circuit} + P_{PA} + P_{dec})T_{on-time-code} + (P_{tx-circuit} + P_{rx-circuit})T_{start} \quad (16)$$

Where :
- $P_{enc}$ represents the power consumption of the Golay encoder at the first sensor node.
- $P_{dec}$ represents the power consumption of the Golay decoder at the destination node.

However, from (14) and (16) the energy consumption per information bit for coded GMSK is written as:

$$E_{\inf bit-GMSK} = \frac{2}{G_{code}\alpha}(1+\beta)\ N_f\ \sigma^2 \ln(1/P_e) G_d + \frac{P_{circuit} T_{on-time-code}}{L} + \frac{2 P_{syn} T_{start}}{L} \qquad (17)$$

The total energy consumption per information bit for GMSK scheme with selected Golay code over different transmission distances is plotted in figure 4.

The setting parameters considered in our simulations are reported in Table 1.

From this plot, we observe that for small distance the uncoded GMSK outperforms the coded system, so using coding techniques is less efficient. But for large distance (d>50 m) using error correcting codes is energy efficient, we can observe about 47% energy saving at d=100.

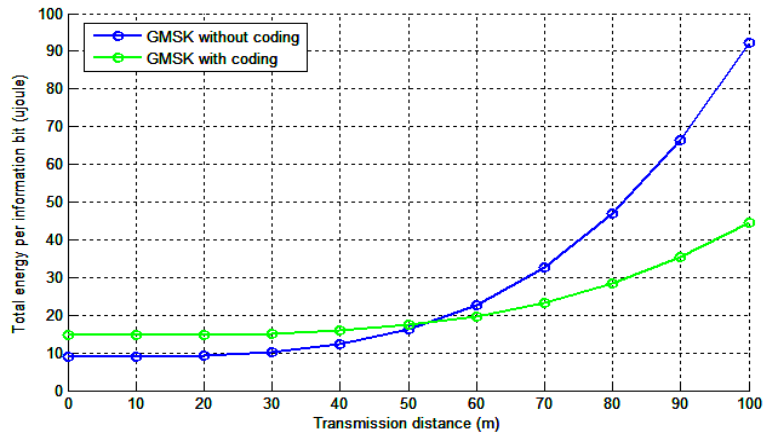

Figure 4. GMSK energy consumption with and without coding technique.

TABLE 1. SIMULATION PARAMETERS

| Parameters | Values |
|---|---|
| $T_{start}$ | $5\ 10^{-6}$ s |
| $T$ | 1.07 s |
| $L$ | $10^3$ bit |
| $\sigma^2$ | $3.981.10^{-21}$ |
| $k$ | 3 |
| $\eta$ | 0.75 |
| $B$ | $10^4$ Hz |
| Carrier frequency | 2.45 GHz |
| $P_e$ | $10^{-4}$ |
| $G_1$ | $10^3$ |
| $M_t$ | $10^4$ |
| $P_{ADC}$ | 6.70 mw |
| $P_{DAC}$ | 15.40 mw |

| Parameters | Values |
|---|---|
| $P_{filt}$ | 2.5 mw |
| $P_{syn}$ | 50 mw |
| $P_{LNA}$ | 20 mw |
| $P_{IFA}$ | 3 mw |
| $P_{mixer}$ | 30.3 mw |
| $P_{enc}$ | 28 mw |
| $P_{dec}$ | 35 mw |
| $G_{code}$ | 4 dB |
| $N_f$ | 10 dB |

## 7. SIMULATION RESULTS

In this study, we examine a small network based on about twenty sensor nodes. These sensors are deployed in non-deterministic mode (randomly) in a 100x100 m$^2$ field as illustrated in figure 6. These nodes are denoted $S_i$ with i= {1,2,…,20}.

Suppose that a source node $S_1$ send L bits of data to the sink. Routing data from the source sensor ($S_1$) node to the destination one can use intermediate sensors to relay information.

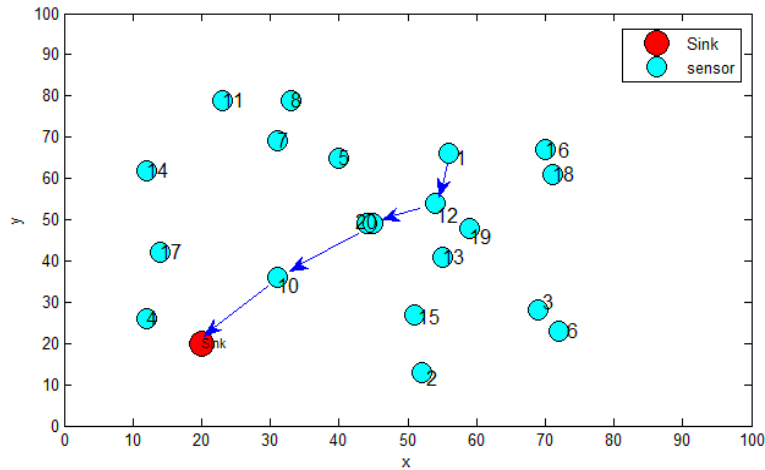

Figure 6. Random distributed wireless sensors network, 100x100 m$^2$

The energy consumed along the route is calculated as follows [20]:

$$E_{Route} = E_{enc} + \sum_{i=1}^{n} P_{tx}(i) T_{on-time-code} + \sum_{i=1}^{n} P_{rx}(i) T_{on-time-code} + E_{dec} \quad (19)$$

Where :
- $P_{tx}$ (i) represents the transmission power of node $S_i$
- $P_{rx}$ (i) represents the receiving power of node $S_i$
- $n$ is the number of node.

25

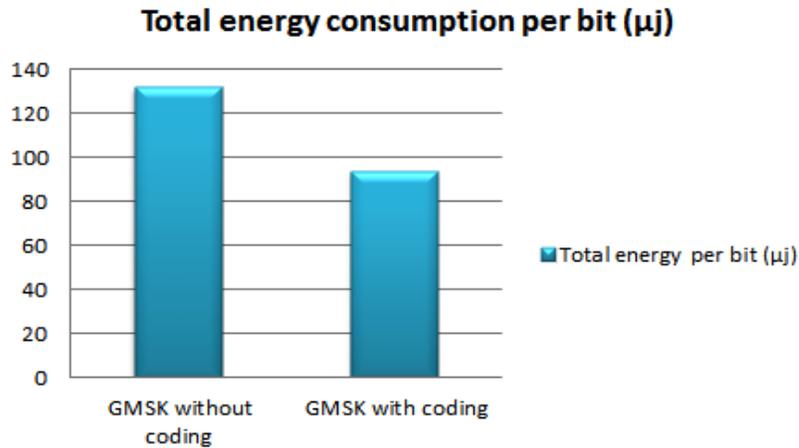

Figure 6. Comparison of energy consumption for both coded and uncoded GMSK

The above graph renders the comparison of total energy consumption during an inter-nodes communication for both uncoded and coded system. In this scenario, the distance between nodes is randomly taken within a range of 50-100m. We assume that there are about 3 intermediate nodes acting as router. The figure 6 reinforces the idea that GMSK scheme with Golay coding process is most energy efficient than the uncoded system. Indeed, we observe that coded system provides about 29% energy saving.

## 8. CONCLUSION

In this paper the performance analysis of GMSK modulation with channel coding approach is studied under AWGN channel condition to enhance reliability of links and reduce the transmitter signal energy in wireless sensor.

Our analytical and simulation results indicate that GMSK scheme with Golay strategy achieves significant gains and that energy consumption can be greatly minimized.

The results also reveal that, GMSK technique may be a good choice for wireless sensor network, because this modulation schemes has the possibility to minimize the sideband power, which permits to decrease the interference between signal carriers. More, It has an excellent power efficiency due to their constant envelope, provides a good BER performance and it has the advantage of being simple to generate and simple to demodulate.

## ACKNOWLEDGEMENTS

This work was supported in part by Laboratory of Acoustics at University of Maine, LAUM UMR CNRS n°6613 in France and Laboratory Innov'COM, University of Carthage, Sup'Com in Tunisia.

Rajoua Anane author wishes to express her sincere gratitude to Professor Ammar Bouallegue for guiding her throughout the current research work.

**Authors**

**Rajoua ANANE** received the B.S. degree in 2008 from Higher Institute of Computer and Communication Techniques, Tunisia and M.S. degree in 2010 from National Engineering School of Tunis. Currently he is a Ph.D. student at the School of Engineering of Tunis. He is a researcher associate with Laboratory of Acoustics at University of Maine,LAUM UMR CNRS n° 6613, France, and Laboratory Innov'Com at Higher School of Communications (SupCom), University of Carthage, Tunisia. 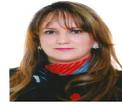

**Ridha Bouallegue** received the Ph.D. degrees in electronic engineering from the National Engineering School of Tunis. In Mars 2003, he received the Hd.R degrees in multiuser detection in wireless communications. From September 1990 He was a graduate Professor in the higher school of communications of Tunis (SUP'COM), he has taught courses in communications and electronics. From 2005 to 2008, he was the Director of the National engineering school of Sousse. In 2006, he was a member of the national committee of science technology. Since 2005, he was the laboratory research in telecommunication Director's at SUP'COM. From 2005, he served as a member of the scientific committee of validation of thesis and Hd.R in the higher engineering school of Tunis. 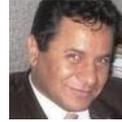

**Kosai Raoof** obtained his M.Sc. and Ph.D. from Grenoble University in 1990 and 1993 respectively; in 1998 he obtained the Habilitation à Diriger Des Recherches Degree (HDR). He was invited to join Laboratoire des Images et Signaux (LIS) in 1999, to participate in the founding of telecommunication research group. His research interest was first focalized on advanced MIMO systems and joint CDMA synchronization; he studied and introduced polarized diversity in MIMO systems. In 2007 he joined GIPSA-LAB to continue his research on Smart Sensor Networks and cooperative MIMO antenna systems. He is a referee for many international journals and conferences in the field of telecommunications and signal processing. He published more than 70 papers and project reports in the field of signal processing. He was granted for European IST Program QTPACK and RNRT French research program ASTURIES in telecommunications in 2004 .He is currently a full professor at the ENSIM Engineering College, University of Maine, Le Mans. 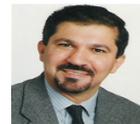